\documentclass[usenatbib]{mn2e}
\input epsf

\title[Periodicity Search of Possible X-ray Counterparts]{Periodicity Search 
of Possible X-ray Counterparts to Radio-quiet Gamma-ray Pulsar Candidates}
\author[L. C.-C. Lin and H.-K. Chang]{Lupin Chun-Che Lin and Hsiang-Kuang Chang\thanks{E-mail:
hkchang@phys.nthu.edu.tw}\\
Department of Physics and Institute of Astronomy, National Tsing Hua university, Hsinchu 30013, Taiwan}

\begin{document}

\date{November 2007; February 2008}
\pagerange{\pageref{firstpage}--\pageref{lastpage}}
\pubyear{2008}
\maketitle

\label{firstpage}

\begin{abstract}
Periodicity search in gamma-ray data is usually difficult 
because of the small number of detected photons. 
A periodicity in the timing signal at other energy bands from 
the counterpart to the gamma-ray source may help to 
establish the periodicity in the gamma-ray emission 
and strengthen the identification of the source in different energy bands. 
It may, however, still be difficult to find the period directly 
from X-ray data because of limited exposure. 
We developed a procedure, by cross-checking two X-ray data sets, 
to find candidate periods for X-ray sources which are possible
counterparts to gamma-ray pulsar candidates.
Here we report the results of this method 
obtained with all the currently available X-ray data of 8 X-ray sources. 
Some tempting periodicity features were found.
Those candidate periods can serve as the target periods 
for future search when new data become available so that
a blind search with a huge number of trials can be avoided. 
\end{abstract}


\section{Introduction} 

The electrodynamics in the magnetosphere of a neutron star
has not yet been well understood, although fascinating phenomena
as revealed in emissions across the whole electromagetic spectrum
from neutron stars have been observed for forty years.
Among the 2000 or so radio pulsars cataloged to date, 
only 7 are detected at photon energies higher than 100 MeV
and 3 more are detected at a lower confidence level \citep{thompson04}.
These 7 include Geminga, which does not emit radio waves or emits only
at a very low level \citep{kuzmin99}.
The effort to understand gamma-ray emissions from pulsars has resulted in
two types of models: the slot-gap models \citep{muslimov04} and the outer-gap models
(e.g. \citet{cheng00,hirotani07,takata07}).
These models, although different in the assumed emission site and mechanisms,
have been successful in many aspects in explaining the pulse profiles and spectra.
However, they do give different estimates of the fraction of radio-quiet gamma-ray pulsars,
like Geminga, among all the gamma-ray pulsars \citep{harding07}.
A gamma-ray pulsar may be radio-quiet if its gamma-ray beam is wider than its radio beam 
or these beams are in different directions. It is also possible that it is
intrinsically weak in radio emission because of unfavored spin periods and magnetic field strength. 
Although a consistent model for radio emission from pulsars is still lacking, several empirical
models are available.
Given a radio emission model (e.g. 
\citet{johnston06}), different models for
gamma-ray pulsars may be discriminated by comparing their estimates of
the radio-quiet gamma-ray pulsar fraction with observations, in addition to comparing
their prediction of high-energy emission polarization characteristics (e.g.\
\citet{takata07b}).

The number of currently known gamma-ray pulsars is only 7, which is obviously too small to
give a meaningful fraction of the radio-quiet ones. 
With the launch of GLAST and AGILE, the number of gamma-ray pulsars is expected to increase significantly.
However, radio-quiet gamma-ray pulsars may still be difficult to find, mainly because of the limited amount
of gamma-ray photons. 
Just like the case of Geminga, only after its period was discovered in the ROSAT data
\citep{halpern92}, its pulsation in gamma-rays was then found. 
Very often it is also difficult to find the period in X-ray data, particularly when the data is sparse and a blind search
over a large range of periods is performed.
Efforts to find new radio-quiet gamma-ray pulsars from unidentified EGRET sources in a multiwavelength approach
have not yet yielded definite results (e.g.\ 
\citet{reimer01,halpern02}). 
Among the unidentified EGRET sources, many are at low galactic latitude and may be gamma-ray pulsars.
Continuing this effort, we conducted a periodicity search for possible X-ray counterparts to some bright unidentified
EGRET sources, using all available archival data with the method of cross-checking two data sets to look for candidate periods. 
The purpose of this cross-checking procedure is to find relatively
significant signatures of possible pulsations occurring in two independent data sets. 
Candidate periods and the corresponding period time derivatives
thus found are reported and can serve as the target
periods to verify in future X-ray or gamma-ray data.

\section{Source selection and data reducton}

As an effort to understand the nature of unidentified EGRET sources,
\citet{roberts01} presented
a catalog of 2-10 keV 
ASCA/GIS images of 
fields containing 30 bright GeV sources. 
These GeV sources were selected from the brightest ones in
the catalog compiled by  
\citet{lamb97}  using only 
photons with energies above 1 GeV from the EGRET public 
archives that incorporate the first 4.5 yr of CGRO/EGRET observations.
Among these GeV sources, some are identified pulsars and quasars,
some are possible counterparts to previously known objects, and
the others are not yet identified to sources in other energy bands.
Possible X-ray counterparts of these unidentified GeV sources were
proposed in 
\citet{roberts01}.
To search for new radio-quiet gamma-ray pulsars,
we look for periodicity of possible X-ray counterparts to those unidentified
GeV sources for which the possibility of being gamma-ray pulsars has not yet been ruled out. 
Furthermore, as explained in the next section, two data sets adequate for timing analysis
are needed for the cross-checking method
to find candidate periods. 
6 GeV sources among the 30 in the catalog of 
\citet{roberts01} were thus selected.
One of the 6 sources (GeV J1417-6100)
has four potential X-ray counterparts. 
Another one of the 6 sources (GeV J0008+7304) 
has been investigated with the same procedure described in this paper.
The periodicity search result of its possible X-ray counterpart,
RX J0007.0+7302, was reported by
\citet{lin05} and will not be repeated here. 
The 8 X-ray sources and their data used in this study are listed in Table~\ref{data}.
Detailed descriptions of each GeV sources can be found in 
\citet{roberts01}. 
\begin{table*}
\caption{List of data employed in this study. 
Column 1 is the name of the X-ray source with its possible gamma-ray counterpart in the parentheses. 
Column 2 is the instrument used to obtain that data. 
Column 3 is the data epoch (in MJD), which is the middle point in the whole data time span.
Column 4 and 5 are the total exposure time  and the total time span of the data used. Both are in units of ks.
Column 6 and 7 are the target position, photon numbers after extraction, the radius of the extraction circle, and
the extraction energy range (in keV) for the X-ray target under study.
Column 8 is the sequence number to label the data used in this work for later referencing.
}
\label{data}
\begin{tabular}{cccccccc}
\hline
\hline
(1) & (2) & (3) & (4) & (5) & (6) & (7) & (8) \\ 
\hline
AX J1418.2--6047 & ASCA/GIS & 50317.00727 & 25.6 & 106 & ($14^h18^m15^s.8,-60^{\circ}46'44''$) &   311; $90''$; 1-10 &  01 \\
(GeV J1417--6100) & ASCA/GIS & 51223.68497 & 22.2 & 173 & ($14^h18^m15^s.8,-60^{\circ}46'44''$) &  126; $90''$; 1-6 &  02  \\
               & BeppoSAX/MECS & 52289.79012 & 30.8 & 51.0 & ($14^h18^m15^s.8,-60^{\circ}46'44''$) &  105; $120''$; 2-10 &  03 \\
\hline
AX J1418.6--6045 & ASCA/GIS & 50317.00727 & 25.6 & 106 & ($14^h18^m37^s.0,-60^{\circ}45'12''$) &  271; $90''$; 1-8 & 04 \\
(GeV J1417--6100) & ASCA/GIS & 51223.68497 & 22.2 & 173 & ($14^h18^m37^s.0,-60^{\circ}45'12''$) &  133; $90''$; 1-7 & 05\\
                & BeppoSAX/MECS & 52289.79013 & 30.8 & 51.0 & ($14^h18^m37^s.0,-60^{\circ}45'12''$) &  148; $120''$; 2-10 & 06 \\
\hline
AX J1418.7--6058 & ASCA/GIS & 50317.00728 & 25.6 & 106 & ($14^h18^m38^s.6,-60^{\circ}57'49''$) &  314; $90''$; 1-8 & 07 \\
(GeV J1417--6100) & ASCA/GIS & 51223.68496 & 22.2 & 173 & ($14^h18^m38^s.6,-60^{\circ}57'49''$) &  551; $90''$; 1-7 & 08\\
                & BeppoSAX/MECS & 52289.79050 & 30.8 & 51.0 & ($14^h18^m38^s.6,-60^{\circ}57'49''$) &  569; $210''$; 2-10 & 09 \\
               & XMM/pn & 52708.65941 & 26.9 & 26.9 & ($14^h18^m42^s.8,-60^{\circ}58'03''.6$) &  632; $15''$; 0.2-12 & 10 \\
\hline
AX J1420.1--6049 & ASCA/GIS & 50317.00729 & 25.6 & 106 & ($14^h20^m07^s.8,-60^{\circ}48'56''$) &  214; $90''$; 1.5-8 & 11 \\
(GeV J1417--6100) & ASCA/GIS & 51223.68495 & 22.2 & 173 & ($14^h20^m07^s.8,-60^{\circ}48'56''$) &  252; $90''$; 1.5-8 & 12\\
                & BeppoSAX/MECS & 52289.79017 & 30.8 & 51.0 & ($14^h20^m07^s.8,-60^{\circ}48'56''$) &  904; $210''$; 2-10 & 13 \\
               & Chandra/ACIS & 52534.12055 & 31.2 & 31.2 & ($14^h20^m08^s.2,-60^{\circ}48'16''.9$) &  229; $2''$; 1-7 & 14 \\
\hline
AX J1809.8--2332 & ASCA/GIS & 50526.78845 & 41.5 & 180 & ($18^h09^m48^s.6,-23^{\circ}32'09''$) &  1133; $90''$; 0.8-7 & 15 \\
(GeV J1809--2327) & XMM/pn & 52539.07915 & 11.1 & 15.6 & ($18^h09^m50^s.0,-23^{\circ}32'24''$) &  542; $15''$; 0.2-12 & 16 \\
              & XMM/pn & 53280.71956 & 49.1 & 69.3 & ($18^h09^m50^s.0,-23^{\circ}32'24''$) &  4070; $15''$; 0.2-12 &  17 \\
\hline
AX J1836.2+5928 & ROSAT/HRI & 50800.00499 & 40.1 & 443 & ($18^h36^m13^s.6,+59^{\circ}25'29''$) &  224; $30''$; 0.1-2.5 &  18 \\
(GeV J1835+5921) &  Chandra/HRC & 53430.07235 & 45.2 & 45.2 & ($18^h36^m13^s.7,+59^{\circ}25'30''$) &  278; $1''$; 0.08-10 &  19 \\
            & Chandra/HRC & 53438.80937 & 28.1 & 28.1 & ($18^h36^m13^s.7,+59^{\circ}25'30''$) &  181; $1''$; 0.08-10 &  20 \\
            & Chandra/HRC & 53440.60818 & 45.2 & 45.2 & ($18^h36^m13^s.7,+59^{\circ}25'30''$) &  247; $1''$; 0.08-10 &  21 \\
\hline
AX J1837.5--0610 & ASCA/GIS & 50905.53135 & 19.7 & 37.0 & ($18^h37^m29^s.0,-06^{\circ}09'38''$) &  243; $90''$; 1-9 &  22 \\
(GeV J1837--0610) & ASCA/GIS & 51105.10478 & 17.9 & 40.0 & ($18^h37^m29^s.0,-06^{\circ}09'38''$) &  203; $90''$; 1-9 &  23 \\
               & BeppoSAX/MECS & 51981.25968 & 40.2 & 94.5 & ($18^h37^m32^s.5,-06^{\circ}09'49''$) &  739; $210''$; 2-10 &  24 \\
               & BeppoSAX/MECS & 52016.65630 & 42.7 & 105 & ($18^h37^m32^s.5,-06^{\circ}09'49''$) &  770; $210''$; 2-10  &  25 \\
\hline
AX J2021.1+3651 & ASCA/GIS & 51300.06891 & 21.8 & 69.8 & ($20^h21^m07^s.8,+36^{\circ}51'19''$) &  294; $90''$; 1-7 &  26 \\
(GeV J2020+3658) & Chandra/ACIS & 52682.60172 & 20.8 & 20.8 & ($20^h21^m05^s.5,+36^{\circ}51'04''.8$) &  575; $1''$; 0.2-5 &  27 \\
\hline
\end{tabular}
\end{table*}

GeV J1417--6100 is in the direction of the `Kookaburra' radio complex
\citep{roberts99}. 
A 68-ms radio pulsar, PSR J1420--6048, was found at a position consistent with 
AX J1420.1--6049 
\citep{damico01}, which is in the northeastern wing of the Kookaburra. 
The pulsation at the radio period in X-rays was only
marginally detected with ASCA data 
\citep{roberts01b}. 
Chandra images of AX J1420.1--6049 clearly show a compact source at the
position of PSR J1420--6048, but its pulsation in X-rays at the radio
period cannot be confirmed with the current Chandra data
\citep{ng05}.
The identification between PSR J1420--6048 and AX J1420.1--6049 
is not yet conclusive.
We therefore include AX J1420.1--6049 in our periodicity search.
AX J1418.7--6058 is located in the southwestern wing of the Kookaburra 
(the `Rabbit'). 
Two compact sources denoted as `R1' and `R2' in the Rabbit
were identified in Chandra and XMM 
images
\citep{ng05}. 
Although `R1' might be a background source, 
we still used it as the target in the XMM data 
for cross-checking periodicity signatures
with earlier ASCA and BeppoSAX data, since `R1' is much brighter than `R2' and
in those earlier data
photons from `R1' dominate. 
More recently, two extended TeV sources (HESS J1420--607 and HESS J1418--609), 
spatially coincident with the two wings of the Kookaburra,
were discovered 
\citep{aharonian06}. 
It is likely that 3EG J1420--6038 and GeV J1417--6100 are two separate sources
associated with HESS J1420--607 and HESS J1418--609 respectively
\citep{aharonian06}.
Another two X-ray sources, AX J1418.2--6047 and AX J1418.6--6045, denoted as
`Src2' and `Src3' in 
\citet{roberts01}, are both located in the 95\% error contour of GeV J1417--6100
and in the 99\% contour of 3EG J1420--6038
\citep{aharonian06}.

AX J1809.8--2332 is the brightest X-ray source within the 99\% error box 
of GeV J1809--2327 (3EG J1809--2328).
Chandra images of this region reveal a point X-ray source connected to a
nonthermal X-ray/radio nebula
\citep{braje02}. 
XMM images show a larger, broader X-ray emission region with a ridge 
of emission along the symmetric axis of the radio nebula
\citep{roberts06}.
The X-ray point source is very likely a pulsar moving 
at the tip of its associated pulsar wind nebula.
It is therefore very much desired to find its pulsation period. 

GeV J1835+5921 and its possible X-ray counterpart AX J1836.2+5928 have
long been considered as the next Geminga to find.
Strong evidence from Chandra and HST observations indicates
AX J1836.2+5928 (RX J1836.2+5925) is a neutron star
\citep{halpern02}.  
Using the Chandra data listed in Table~\ref{data}, 
\citet{halpern07} reported 
unsuccessful searches for pulsations from RX J1836.2+5925 
in the period range from 1 ms to 10 s
and a pulsed-fraction upper limit 
of 35\%. 

There are fewer observations of the GeV J1837--0610 region.
A 96-ms radio pulsar (PSR J1837--0604) was discovered and suggested to be
associated with GeV J1837--0610 (3EG J1837--0606)
\citep{damico01}.
However, the position of PSR J1837--0604 is at the rim of the 95\% confidence
contour of GeV J1837--0610 and is inconsistent with AX J1837.5--0610, 
which is the single X-ray point source within the 95\% confidence contour of 
GeV J1837--0610 in ASCA images.  

Chandra observations of AX J2021.1+3651 
\citep{hessels04} indicate that
this X-ray source is a pulsar wind nebula and there is an X-ray
point source located at the center of this nebula, where 
a 104-ms radio pulsar (PSR J2021+3651) was found
\citep{roberts02}. 
The X-ray pulsation of this point source at the radio period was
only marginally found in the Chandra data with an $H$-test score of 11.7
\citep{hessels04}. 

All the data were processed with standard procedures as outlined below. 
Photons for timing analysis were selected within a circle around the target
and within a certain energy range. The circle radii and energy-range boundaries 
were so chosen that the target contribution was clearly detected.
The target positions, the radii of the circles and the energy ranges adopted in this work are listed
in Table~\ref{data}. 
All the photon arrival times were then corrected to the solar system barycenter.
The corrected time list was then used for timing analysis to look for periodicity.

\section{The cross-checking method for periodicity search}

Periodicity search in sparse data usually employs photon-counting
coherent-timing methods, that is, each photon arrival time is converted to
a phase value with a given period and its time derivatives. The distribution of these phase values
is then tested against the null hypothesis with a certain statistical method
to assess the probability that the observed distribution is consistent with
a statistically flat distribution.
In this work, we adopted the $H$-test 
\citep{dejager89,dejager94} as the method to assess
the random probability.

The difficulty to find the period in sparse data, in addition to the limited 
number of observations, 
is that a blind search with a huge number of trial periods (and period time derivatives)
is unavoidable if an a priori period or period range is not known. 
When the total number of independent trials is taken into account, one seldom obtains
a significantly low random probability. 
A possible way to improve upon this is to cross-check two different data to look for
possible support for the reality of a tempting candidate period found in one data set.

\begin{figure}
\epsfxsize=8.4cm
\epsffile{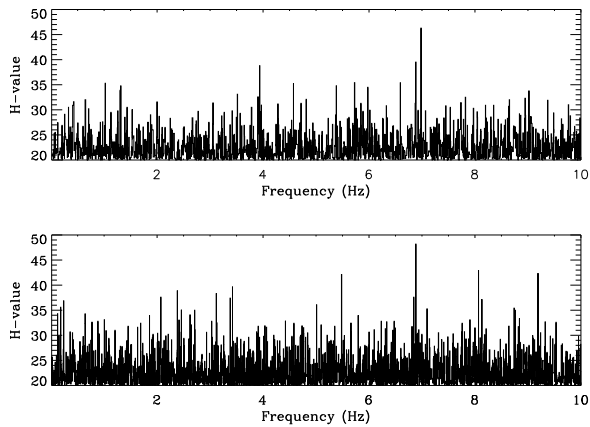}
\caption{Examples of $H$-test blind search results. 
Shown here are those of AX J1420.1--6049 with data sets 11 and 12 (see Table~\ref{data}). 
The number of independent trial frequencies is about $10^6$ for each case.
Most of the trials yield a very low $H$-value. 
Only $H$-values larger than 20 are plotted.  }
\label{blind}
\end{figure}

\begin{figure}
\epsfxsize=8.4cm
\epsffile{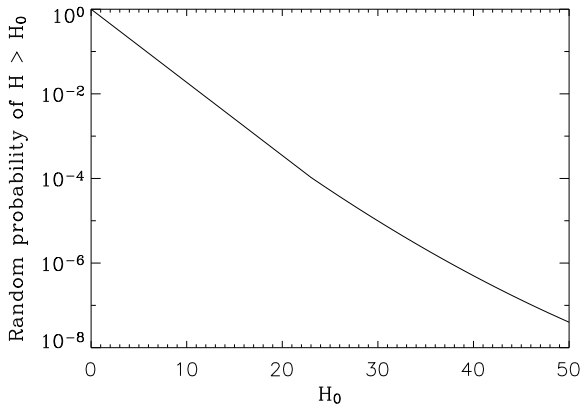}
\caption{The $H$-test probability distribution \citep{dejager89}. 
The plotted curve is the null hypothesis random probability for a single trial to have an $H$-value larger
than $H_0$.}
\label{htest}
\end{figure}

For each of the X-ray sources discussed in the previous section 
we first performed a blind search in every available data of that source.
Depending on the time resolution of the data 
different search ranges of trial periods were chosen. 
For all the ASCA/GIS data, the searching range is from
0.1 s to 100 s. 
For all the others, it is from 0.01 s to 100 s.
Since each data set spans a relatively short time,
we constrained ourselves at this stage to consider trial periods only 
without their time derivatives.
This makes the search a one-dimensional one.
We chose the searching step to be a small fraction (typically 1/5) of the corresponding independent Fourier spacing,
which is $P^2/T$ in the period domain ($1/T$ in frequency),
where $P$ is the trial period and $T$ is the total time span of the data.
A blind search usually gives several tempting features. 
One example is shown in Fig.~\ref{blind}. 
For readers' convenience, the random probability for a single trial to have an $H$-value larger
than a certain number is plotted in
Fig.~\ref{htest}.

\begin{figure}
\epsfxsize=8.4cm
\epsffile{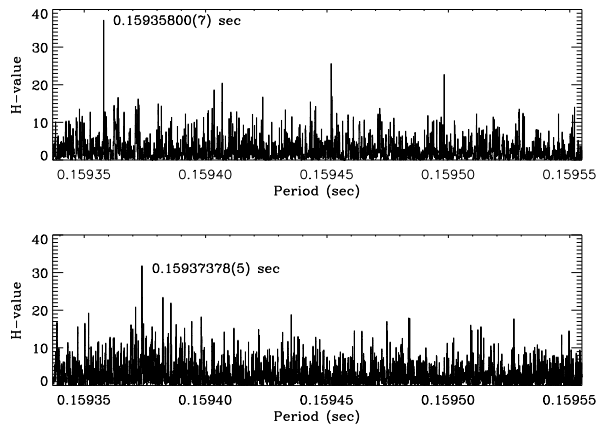}
\caption{The identification of the candidate period at 159 ms for AX J1418.6--6045 using data sets 04 and 05.
The trial period range shown in the lower panel of this figure 
(from data set 05) is roughly the range specified by the criterion that the
corresponding characteristic age is larger than 1000 years 
if a possible period is to be associated with the one in the upper panel.
The most significant feature in that range is at 0.15937378 sec, and its
association with the one at 0.15935800 sec in the upper panel 
yields a resultant random probability of $5.3\times 10^{-3}$. 
We list all such cases with a random probability lower than 0.01 as
a candidate period in Table~\ref{pcan}.   }
\label{asso1}
\end{figure}

\begin{figure}
\epsfxsize=8.4cm
\epsffile{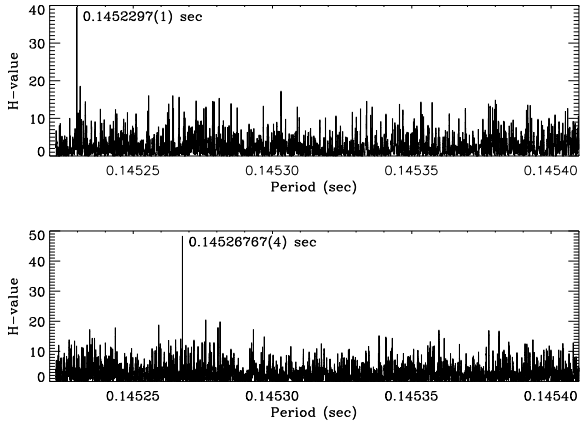}
\caption{The identification of the candidate period at 145 ms for AX J1420.1--6049 using data sets 11 and 12.
The same as in Figure~\ref{asso1}, 
the trial period range shown in this figure is roughly the range specified by the criterion that the
characteristic age is larger than 1000 years. This figure is in fact a zoom-in view of Figure~\ref{blind} around 
6.88564 Hz but plotted in the period domain.
This example, with a resultant random probability of $4.1\times 10^{-5}$, is 
the most significant signature found in this study.}
\label{asso2}
\end{figure}

From the result of a blind search in a certain data set, 
a few trial periods with high $H$-values were picked out
for cross-checking. The number of such trial periods to pick out is obviously arbitrary. 
We picked out the top five in this work. 
Each of these five usually has a significantly low random probability if the total
number of independent trials in that blind search is not taken into account. 
As mentioned above, a blind search has a huge number of independent trials, 
typically more than $10^6$. Therefore, for each of them, we look for possible associated
features in other data sets to assess its reality.
To do this, we further set two criteria.
At first we look for associated features of the $H$-test results in another data set 
only within the range that gives
a corresponding characteristic age ($\frac{P}{2\dot{P}}$) 
larger than 1000 years.
A neutron star younger than that age probably would have a record in human history.
In determining that range, the uncertainty of the period is estimated based on the method described in
\citet{leahy87}.
Secondly, the most prominent features in that range are identified and the resultant random probability
is calculated, which is the product of the random probability of the feature in the first data set and
that of the feature in the second data set. When calculating the random probability of the feature in the
second data set, only independent trials within the range specified by the first criterion are considered.
Oversampling was not taken into account in our computation of the random probability. 
We keep only those cases whose resultant random probability is less than 0.01. 
The periods and the corresponding period time derivatives for those cases are listed in Table~\ref{pcan}.
The probability threshold of 0.01 is arbitrary. Of course a higher one gives more candidate periods,
but with less significance.
Two examples of such an association are shown in Fig.~\ref{asso1} and \ref{asso2}. 

\begin{table*}
\caption{Candidate periods. 
Column 1 is the candidate period (in seconds) obtained from cross-checking the two data sets listed in column 7.
The number in the parentheses is the 1-$\sigma$ uncertainty to the last digit,
 estimated with the method described in \citet{leahy87}.
Those followed by a triangle $(\triangle)$ 
are candidate periods favoured by examing the consistency with
properties of currently known gamma-ray pulsars, as discussed in Section 4. 
The epoch (MJD) of this period is listed in column 5, which is the epoch of the data set containing the stronger signature for pulsation
in the two data sets.
Column 2, 3 and 4 are the corresponding period time derivative (in units of $10^{-13}$ s s$^{-1}$), 
characteristic age (in years) and spin-down power (in units of $10^{36}$ erg s$^{-1}$).
Column 6 is the resultant random probability as explained in the text. 
}
\begin{tabular}{lcccccc}
\hline
\hline
(1) & (2) & (3) & (4) & (5) & (6) & (7) \\
\hline
 AX J1418.2--6047  &  &  &  &  &  &  \\  
\hline
0.2029690(1) $\triangle$ & 2.00(4) & 16000 &  9.4(2) & 50317.00727 & 4.7$\times 10^{-3}$ & 01, 03 \\
0.3097181(5) & 47.3(1) & 1000 &  6.29(1) & 51223.68497 & 2.7$\times 10^{-3}$ & 01, 02 \\
0.4289840(5) & 33.6(1) & 2000 &  1.682(6) & 51223.68497 & 5.3$\times 10^{-3}$ & 01, 02\\
\hline
 AX J1418.6--6045 &  &  &  &  &  &  \\   
\hline
0.12180376(5) & 6.65(2) & 2900 &  14.51(4) & 51223.68497 & 9.1$\times 10^{-3}$ & 05, 06 \\
0.15935800(7) $\triangle$ & 2.01(1) & 12500 &  1.96(1) & 50317.00727 & 5.3$\times 10^{-3}$ & 04, 05 \\
0.3311744(5) & 39.39(9) & 1300 &  4.275(9) & 50317.00727 & 2.7$\times 10^{-3}$ & 04, 05 \\
0.3311744(5) & 31.04(6) & 1700 &  3.366(7) & 50317.00727 & 3.2$\times 10^{-4}$ & 04, 06 \\
\hline
 AX J1418.7--6058 &  &  &  &  &  &  \\   
\hline 
0.10139550(2) & 0.080(4) & 200000 &  0.30(2) & 50317.00728 & 9.6$\times 10^{-3}$ & 07, 09 \\
0.10139550(2) & 7.757(5) & 2100 &  29.32(2) & 50317.00728 & 3.6$\times 10^{-3}$ & 07, 09 \\
0.10171064(5) &  9.337(6)  &  1700  & 34.95(2) & 50317.00728 & $< 3.9\times 10^{-3}$ & 07, 09 \\
0.10969281(4) & 14.45(1) & 1200 &  43.03(3) & 50317.00728 & 4.2$\times 10^{-3}$ & 07, 10  \\
0.1443257(1) & 18.35(2) & 1250 &  24.13(2) & 51223.68496 & 1.9$\times 10^{-4}$ & 07, 08 \\
0.3274125(3) & 31.29(8) & 1650 &  3.515(9) & 51223.68496 & 9.5$\times 10^{-3}$ & 08, 09 \\
\hline
 AX J1420.1--6049 &  &  &  &  &  &  \\ 
\hline 
0.014799899(6) & 0.392(3) & 6000 &  478(4) & 52534.12055 & $< 6.2\times 10^{-3}$ & 13, 14 \\
0.019576788(3) & 2.616(3) & 1200 &  1376(1) & 52289.79017 & 7.2$\times 10^{-3}$ & 13, 14 \\
0.033549387(8) & 0.133(7) & 40000 &  13.9(8) & 52289.79017 & 1.2$\times 10^{-3}$ & 13, 14 \\
0.14526767(4) & 4.84(2) & 4750 &  6.24(2) & 51223.68495  & 4.1$\times 10^{-5}$ & 11, 12 \\
0.14526767(4) $\triangle$ & 1.35(3) & 17000 &  1.74(3) & 51223.68495 & 5.0$\times 10^{-3}$ & 12, 13 \\ 
0.1881326(4) & 10.1(3) & 2900 &  6.0(2) & 52534.12055 & 3.8$\times 10^{-3}$ & 13, 14\\
0.2540350(5) & 13.17(7) & 3100 &  3.17(2) & 50317.00729 & 5.0$\times 10^{-3}$ & 11, 12 \\
\hline
 AX J1809.8--2332 &  &  &  &  &  &  \\  
\hline 
0.13247455(3) & 16.439(5) & 1300 & 27.792(8) & 50526.78845 & $< 1.9\times 10^{-2}$ & 15, 17 \\
0.2649491(2) & 27.4(1) & 1500 &  5.80(2) & 50526.78845 & 7.4$\times 10^{-4}$ & 15, 16 \\
\hline
 AX J1836.2+5928 &  &  &  &  &  &  \\   
\hline 
0.022425706(5) & 0.77(9) & 5000 &  2.7(3) $\times 10^{2}$ & 53440.60818 & 6.5$\times 10^{-3}$ & 19, 21 \\
0.03350910(2) & 2(2) & 2000 &  2(2) $\times 10^{2}$  & 53438.80937 & 5.3$\times 10^{-3}$ & 20, 21 \\
0.034677614(1) & 0.0201(4) & 270000 &  1.90(4) & 50800.00499 & 5.5$\times 10^{-3}$ & 18, 19 \\
0.1679234(4) $\triangle$ & 0.16(2) & 170000 &  0.13(1) & 53440.60818 & 9.1$\times 10^{-3}$ & 18, 21 \\
\hline
 AX J1837.5--0610 &  &  &  &  &  &  \\ 
\hline 
0.1736610(3) & 18.6(5) & 1500 &  14.0(3) & 51105.10478 & 2.8$\times 10^{-3}$ & 22, 23 \\
\hline
 AX J2021.1+3651 &  &  &  &  &  &  \\ 
\hline 
none & & & & & & \\
\hline
\end{tabular}
\label{pcan}
\end{table*}

\begin{figure}
\epsfxsize=8.4cm
\epsffile{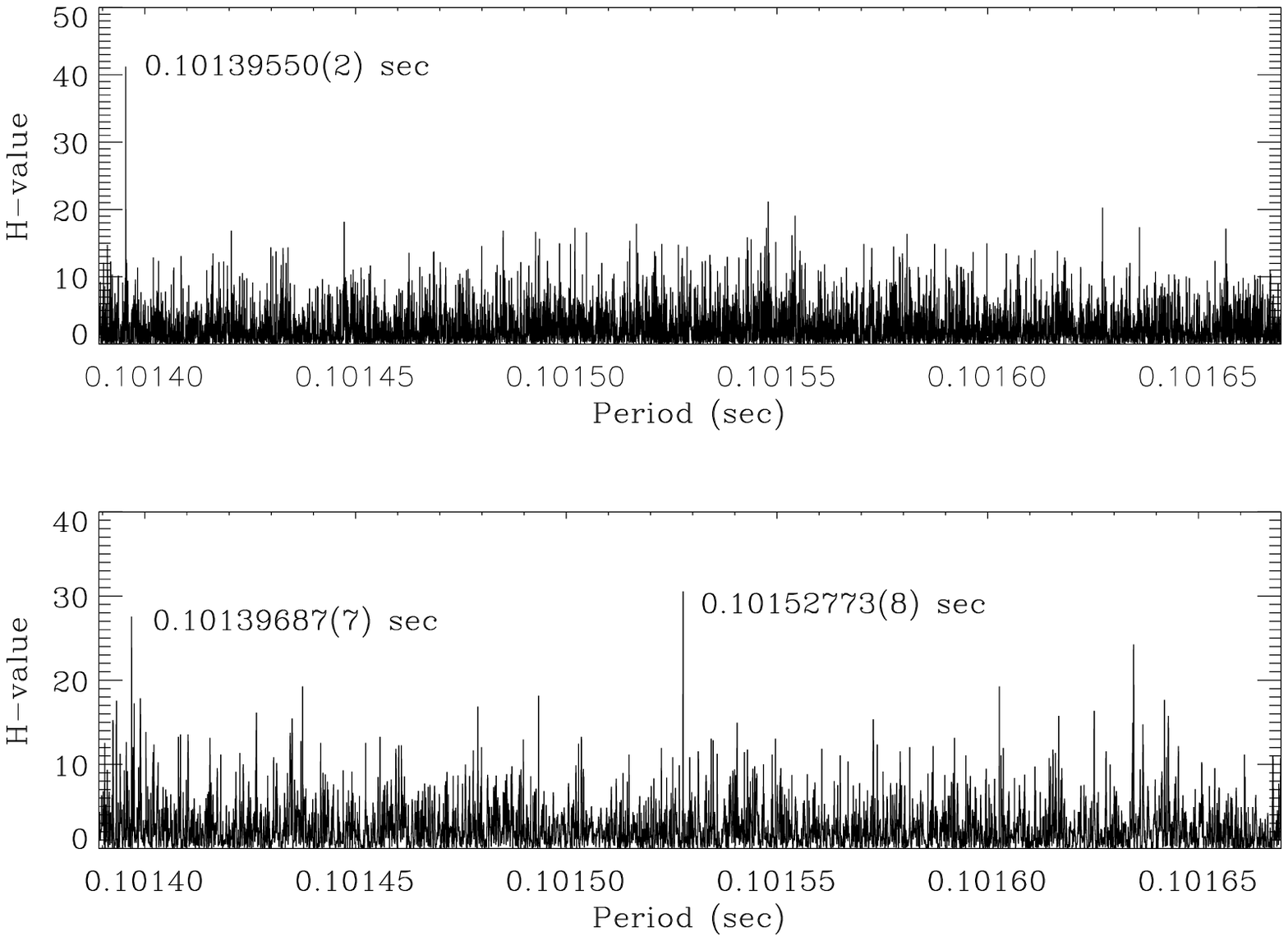}
\caption{The identification of the candidate period at 101.4 ms for AX J1418.7--6058 using data sets 07 and 09.
The same as in Figure~\ref{asso1} and \ref{asso2}, 
the trial period range shown in the lower panel of this  
figure is roughly the range specified by the criterion that the
characteristic age is larger than 1000 years. 
The two possible period associations, which result in different period time derivatives, are both included in Table~\ref{pcan}. 
}
\label{asso3}
\end{figure}

\begin{figure}
\epsfxsize=8.4cm
\epsffile{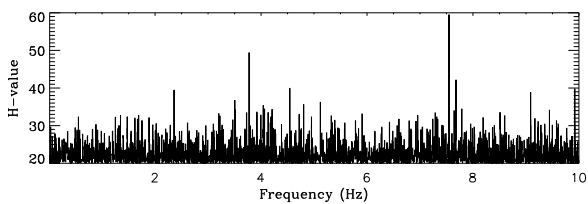}
\caption{The $H$-test blind search results of AX J1809.8--2332 
with data set 15. 
The two highest features correspond to periods at 0.13247455 s and
0.2649491 s respectively.
}
\label{hdis1809}
\end{figure}

In data set 07, the candidate period of AX J1418.7--6058 at 0.10171064 s has an $H$-value of 56.7, which has a
very low random probability for a single trial. 
With our procedure, there will be candidate periods satisfying our criteria in all the data sets 08, 09, and 10.
Here we only report the most significant one. 
Its random probability listed in Table \ref{pcan} was derived using an
$H$-value equal to 50 instead of 56.7, since in \citet{dejager89} 
the signle trial random probability was provided only up to $H$ equal to 50. 
Two similar cases are the candidate period at 0.014799899 s 
for AX J1420.1--6049, which has an $H$-value of 53.8 in data set 14,
and that at 0.13247455 s for AX J1809.8--2332, 
which has an $H$-value of 59.5 in data set 15.
Although in our procedure only the most prominent feature 
in the cross-checking window is picked out for association, for
the candidate period of AX J1418.7--6058 at 0.10139550 sec in data set 07,
the association with the period of the second highest 
$H$-value in the cross-checking window of data set 09
is also reported here, since it is very close to the candidate period 
in data set 07 and its $H$-value is not too different from the highest.
That association is shown in Figure~\ref{asso3}. 
The two candidate periods of AX J1809.8--2332 listed in Table \ref{pcan} are
related to each other. 
Based on their pulse profiles
we suggest the period at 0.13247455 s be the second harmonic of the
one at 0.264949 s 
if they are real.  
In such a case, the derived period derivative, characteristic age and 
spin-down power for the association between periodicity signatures found in
data sets 15 and 17 near 0.1325 s will be 2, 1 and 0.25 times those listed in
Table \ref{pcan} respectively. It makes this association even more unfavoured
if one compares its spin-down power with that of the Crab pulsar; 
see discussion in the next section.   
On the other hand, the strong periodicity signature (Figure~\ref{hdis1809})
makes the 
candidate period at 0.264949 s very tempting.

To examine whether the procedure described above is useful, we applied it to the case of
RX J1856.5--3754, whose X-ray emission is believed to be thermal emission from 
the surface of a neutron star.
There were several XMM observations of that source, but no periodicity was found until
a period of 7.055 s with a random probability of $6\times 10^{-4}$ was reported by 
\citet{tiengo07}. 
That result was obtained by analysing the XMM data taken in October 2006.
Applying our cross-checking method as described above to previous data, 
we found that the 7-s period was also picked out 
and suggested as a candidate period. 
It demonstrates, although not guarantees, the usefulness of our cross-checking method.
  
\section{Discussion}

We applied the periodicity cross-checking procedure described in
Section 3 to the 8 X-ray sources listed in Table~\ref{data}. 
Candidate periods obtained with this method are listed in Table~\ref{pcan}.
Future periodicity searches at X-rays or gamma-rays 
may be conducted around these
candidate periods, instead of performing blind searches. 
Among all these candidate periods, the one with the lowest random probability (4.1$\times10^{-5}$) is
145.2297 ms for AX J1420.1-6049. Its implied characteristic age, 4800 years, is between that of the
Crab pulsar and of the Vela pulsar. However, its corresponding spin-down power is similar to that
of the Vela pulsar, which in turn is about 100 times smaller than that 
of the Crab.
If there is a correlation
between the characteristic age and the spin-down power,
this candidate period is not favoured because
its associated spin-down power is too small for its
4800-year characteristic age. 

Indeed, among the currently known gamma-ray pulsars,
the spin-down power decreases with increasing characteristic ages. 
The Crab pulsar, with a characteristic age of $10^{3.1}$ years, has a spin-down power about
$10^{38.7}$ erg/s. That of the Vela pulsar are $10^{4.1}$ years and $10^{36.8}$ erg/s and 
of Geminga are $10^{5.5}$ years and $10^{34.5}$ erg/s. 
One possible way to further assess the reality of these candidate periods, therefore,
is to compare their corresponding 
properties, such as the characteristic age and the spin-down power, with that of
known gamma-ray pulsars.
Along this line of thinking, the candidate periods 
at 203 ms for AX J1418.2-6047, 
at 159 ms for AX J1418.6-6045, 
at 19.6 ms and 145 ms for AX J1420.1-6049,
and at 22.4 ms, 33.5 ms and 168 ms for AX J1836.2+5928
are favoured, since their corresponding characteristic ages and spin-down powers
are roughly consistent with that of known gamma-ray pulsars.

Another property to consider is the X-ray to gamma-ray energy spectral index, defined in
\citet{roberts01}, which assumes an X-ray counterpart to those unidentified GeV sources
and leads to three X-ray brightness categories for the GeV sources under study. 
In that scheme, the Crab pulsar is X-ray bright, the Vela pulsar is X-ray moderate, and
Geminga is X-ray faint. 
The GeV sources associated with the aforementioned X-ray sources 
are all X-ray moderate (similar to Vela), 
except for GeV J1835+5921, which is X-ray faint (similar to Geminga).
Examing the characteristic ages, we note that the four candidate periods 
at 203 ms for AX J1418.2-6047, 
at 159 ms for AX J1418.6-6045, 
at 145 ms for AX J1420.1-6049
and at 168 ms for AX J1836.2+5928
survive the comparison.
The candidate period at 127.5 ms for RX J0007.0+7302 reported
by \citet{lin05} is also supported by these comparisons. 

These five X-ray sources are possible counterparts to three (or four) 
GeV sources,
namely, GeV J0008+7304, GeV J1417--6100 (or a probable separate source
3EG J1420--6038), and GeV J1835+5921.
For the former two (or three), 
if the candidate periods suggested here turn out to be true,
they will be more Vela-like, in terms of timing properties.
It would suggest that the major cause for a gamma-ray pulsar 
to be radio quiet is
likely geometrical, rather than intrinsic, and the
gamma-ray beam is wider than the radio one so that 
for a certain viewing angle both
the radio and gamma-ray emissions are observed, as for the case of Vela,
and in another viewing angle only the gamma-ray emission is observed.
Instead, if all the radio-quiet gamma-ray pulsars are similar to Geminga 
in their timing properties, the cause of being radio quiet, although not yet
clear, is probably intrinsic.      

Timing properties of not-yet-discovered gamma-ray pulsars may not be similar to
the known ones, particularly because the number of the latter is still small. 
All the candidate periods listed in Table~\ref{pcan} 
may therefore still be helpful 
for pinning down the periodicity of emissions from
these sources in the future when new X-ray data or gamma-ray data obtained with AGILE and GLAST
become available. A larger sample of gamma-ray pulsars, either radio loud or quiet, will help us to understand
better the radiation mechanisms in the magnetospheres of gamma-ray pulsars. 
 
\section*{Acknowledgments}

This research has made use of data obtained through 
the High Energy Astrophysics Science Archive Research Center Online Service, 
provided by the NASA/Goddard Space Flight Center.
This work was supported by the National Science Council and the 
National Space Organization of 
the Republic of China 
through grants NSC 96-2628-M-007-012-MY3 and 
96-NSPO(B)-SP-FA04-01.

\label{lastpage}
\end{document}